\documentclass[12pt,preprint]{aastex}
\newcommand{\bea}{\begin{eqnarray}}
\newcommand{\eea}{\end{eqnarray}}

\def\fun#1#2{\lower3.6pt\vbox{\baselineskip0pt\lineskip.9pt
\ialign{$\mathsurround=0pt#1\hfil##\hfil$\crcr#2\crcr\sim\crcr}}}
\def\lap{\mathrel{\mathpalette\fun <}}
\def\gap{\mathrel{\mathpalette\fun >}}

\def\kms{km s$^{-1}$}

\def\mh{m_\bullet}

\begin{document}
\title{Consequences of Triaxiality for Gravitational Wave Recoil of black holes}
\author{Alessandro Vicari,\altaffilmark{1}
Roberto Capuzzo-Dolcetta,\altaffilmark{1}
David Merritt\altaffilmark{2}}
\altaffiltext{1}{Dipartimento di Fisica, Universita' di Roma La Sapienza
 P.le A.Moro 5, I-00185, Roma, Italy;}
\altaffiltext{2}{Department of Physics, Rochester Institute of Technology,
85 Lomb Memorial Drive, Rochester, NY
14623, USA}

\begin{abstract}
Coalescing binary black holes experience a ``kick'' due to
anisotropic emission of gravitational waves with an amplitude
as great as $\sim 200$ km s$^{-1}$.
We examine the orbital evolution of black holes that have been
kicked from the centers of triaxial galaxies.
Time scales for orbital decay are generally longer in triaxial
galaxies than in equivalent spherical galaxies, since
a kicked black hole does not return directly through the
dense center where the dynamical friction force is highest.
We evaluate this effect by constructing self-consistent triaxial
models and integrating the trajectories of massive particles
after they are ejected from the center;
the dynamical friction force is computed directly from the velocity
dispersion tensor of the self-consistent model.
We find return times that are several times longer than in
a spherical galaxy with the same radial density profile,
particularly in galaxy models with dense centers, implying
a substantially greater probability of finding an off-center
black hole.
\end{abstract}


\section{Introduction}

The coalescence of binary black holes (BHs) results in a gravitational
recoil, or ``kick'', due to the net linear momentum carried
away by gravitational waves.
Bekenstein (1973) estimated a kick velocity $V\approx 300$ \kms\
in highly nonspherical collapses using a quasi-Newtonian formalism.
Nakamura et al. (1987) computed $V/c=0.045\eta^2$ for head-on
collisions from infinity using BH perturbation theory;
here $\eta\equiv\mu/(M_1+M_2)$,
the ``reduced mass ratio'', with $M_1, M_2$ the BH
masses and $\mu\equiv M_1M_2/(M_1+M_2)$ the reduced mass.
A number of analytic estimates have been made of $V$ for
circular-orbit inspirals (Fitchett 1983; Fitchett \& Detweiler 1984;
Wiseman 1992; Favata, Hughes \& Holz 2004).
The kick amplitude is usually divided into two components:
the net recoil up to the
innermost stable circular orbit (ISCO), and the contribution
from the final plunge, from the ISCO to the horizon,
which takes place in the strong-field regime and which
dominates the total kick.
Blanchet, Qusailah \& Will (2005) computed $V$
to second post-Newtonian order and found it to be well
approximated by the simple formula
\begin{equation}
\frac{V} {c} = 0.043\eta^2\sqrt{1-4\eta}\left(1+{ \frac{\eta}{4} }\right).
\label{eq:blanchet}
\end{equation}
The $\eta^2(1-4\eta)^{1/2}$ dependence is the same
found by Fitchett (1983),
who computed gravitational recoil for a pair
of BHs interacting via Newtonian forces and included
only the lowest gravitational wave multipoles
needed for momentum ejection.
With the additional, ad hoc factor in equation~(\ref{eq:blanchet}),
Blanchet, Qusailah \& Will (2005) were able to reproduce the
results of their 2PN calculations
to better than $1\%$ at all mass ratios.
The maximum estimated kick
velocity was $250 \pm 50$ km s$^{-1}$.
Damour \& Gopakumar (2006) estimated a much lower value,
$\sim 74$ km s$^{-1}$, for the maximum kick using an effective
one-body approach.
In the last year, remarkable progress has been made
in techniques for the numerical solution of the full field equations
(Pretorius 2005, 2006; Campanelli et al. 2005;
Campanelli, Kelly \& Lousto 2006; Baker et al. 2006a,b)
allowing several groups to compute recoil velocities for coalescing
BH binaries without approximations.
Baker et al. (2006c) find $V=105\pm 10$ \kms\ for $M_2/M_1=2/3$.
Herrmann, Shoemaker \& Laguna (2006) derive
$V=33$ \kms\ for $M_2/M_1=0.85$ and $V=9$ \kms\ for $M_2/M_1=0.96$.
Most recently, Gonz\'alez et al. (2006) carried out a
large set of inspiral simulations for non-spinning,
circular-orbit binaries
and determined the kick velocity as a function of mass
ratio.
Their results are well described by the expression
\begin{equation}
\frac{V}{c} = 0.040\eta^2\sqrt{1-4\eta}\left(1-0.93\eta\right),
\label{eq:gonzalez}
\end{equation}
implying a maximum kick velocity of $175.7$ \kms\ at
$M_2/M_1=0.36$, somewhat smaller than implied by
equation~(\ref{eq:blanchet}).
The dependence of the kick velocity on orbital eccentricity
was investigated by Sopuerta, Yunes and Laguna (2006)
using the close-limit approximation,
which models the late stages of coalescence as a single
perturbed BH, coupled with a post-Newtonian estimate
of the recoil during the early evolution.
Their results are consistent with a $(1+e)$ dependence of kick
velocity on eccentricity for the low ($e\le 0.1$)
eccentricities which they investigated,
i.e. $V\approx 195$ \kms\ for $e=0.1$.

All of these results were obtained for non-spinning holes.
In the presence of spin, kick velocities might be larger 
and would be nonzero even for $M_1=M_2$ 
(Redmount \& Rees 1989; Favata et al. 2004).
Calculations of the coalescence of spinning BHs are currently
underway (Campanelli, Lousto and Zlochower 2006a, b).
Recent results (Hermann et al. 2007; Campanelli et al. 2007; Koppitz et al. 2007)
actually show how recoil velocity after coalescence of spinning BHs may go up to
$\approx 450$ \kms, reopening the possibility that a merged binary can be
ejected even from the nucleus of a massive host galaxy.

In this paper, we consider some of the consequences of the kicks
for supermassive BHs in galactic nuclei.
A kick velocity of $200$ \kms\ is sufficient to remove
a coalesced BH from a dwarf elliptical galaxy, even if the
latter is embedded in a dark-matter halo (Merritt et al. 2004).
Escape velocities at the centers of luminous elliptical galaxies
are generally greater than $\sim 400$ \kms\ however,
and kicks of the magnitude so far calculated by numerical
relativists would never be expected to remove BHs from such galaxies.
But the kicks could still displace the BHs temporarily from their central
locations, implying a finite probability of finding an
off-center BH in a giant galaxy.
The kicks would also generate long-lived changes
in the central structure of galaxies as the displaced BHs transfer
their orbital energy to the stars via dynamical friction
(Merritt et al. 2004; Boylan-Kolchin et al. 2004; Madau \& Quataert 2004).
The displacements and their side-effects would have
been greater at earlier times, when the gravitational potential wells 
associated with galaxies were shallower
(Volonteri et al. 2003; Merritt et al. 2004; Madau \& Quataert 2004;
Haiman 2004; Yoo \& Miralda-Escude 2004; Libeskind et al. 2006).

In a non-spherical galaxy, an ejected BH does not pass
precisely through the dense center as it falls back,
reducing the mean value of the dynamical friction force
compared with a spherical galaxy.
The result is a more extended period of displacement compared
with estimates based on spherical galaxy models.
Here, we evaluate the effect of nonspherical galaxy geometries on 
BH infall times using fully self-consistent triaxial models.
The models are constructed via orbital superposition as in
Merritt \& Fridman (1996), and the quantities that define the
velocity ellipsoid at every point on the solution grid are
computed and stored as in Merritt (1980).
Given this information, it is possible to compute accurate
estimates of the dynamical friction force that would act
on a massive object, using the expressions in 
Pesce, Capuzzo-Dolcetta \& Vietri (1992) and Capuzzo-Dolcetta
\& Vicari (2006).
The frictional forces are added to the conservative forces from
the triaxial mass distribution when integrating the BH
trajectories.

Among the new effects we find here is an increase in the effective
``escape velocity'' from the center of the galaxy, 
since some of the BH's initial kinetic energy is lost to 
dynamical friction.
The kick velocity needed to escape is also dependent on the
direction of the kick since the frictional force is direction-dependent.
In general, we find that return times are longer in the triaxial geometry
by factors of $\sim $ a few compared with return times in spherical
galaxies having the same, mean radial density profile, and this
translates into a substantially higher probability of finding
a displaced BH.

\section{Method}

A BH ejected from the center of a galaxy moves in response to the
conservative force corresponding to the gravitational
potential $\Phi ({\bf r})$ of the stars
and to the dynamical friction force per unit mass ${\bf f}_{df}({\bf r})$.
Its motion can be approximated by the solution of the differential
equation
\begin{equation}
{\ddot{\bf r}} = -\nabla \Phi + {\bf f}_{df}
\label{eqdiffbase}
\end{equation}
subject to the proper initial conditions.
We assume throughout that neither $\Phi$ nor
$\mathbf{f}_{df}$ depend explicitly on time.

As usual, we transform this second-order differential equation
into a system of first-order equations:
\begin{mathletters}
\begin{eqnarray}
\mathbf{\dot{r}}&=&\mathbf{v}, \\
\mathbf{\dot{v}}&=&-\nabla \Phi+\mathbf{f}_{df} .
\end{eqnarray}
\label{eq:diffeq}
\end{mathletters}

As a consequence of the dynamical friction term,
the orbital energy, $E$, is no longer a conserved
quantity.
Instead we can write
\begin{equation}
\dot E\equiv \dot E_{df}=\mathbf{f}_{df}\cdot \mathbf{v}
\end{equation}
This expression allows us to verify conservation of the total energy
${\cal E} = K+\Phi+E_{df}$,
where $K$ is the kinetic energy and $E_{df}$ the work done
by the dynamical friction force along the trajectory.

We solved the system of differential equations (\ref{eq:diffeq})
numerically, using the 7/8 order Runge-Kutta algorithm of Fehlberg (1968).
\footnote{The FORTRAN version of this algorithm was kindly
made available to us by S. Udry.}

\subsection{Mass model}

We adopted as galaxy models the triaxial generalizations
of the spherical Dehnen (1993) models.
The mass density is
\begin{equation}
\rho (\mathbf r)=\frac{(3-\gamma)M}{4 \pi abc} \frac{1}{m^{\gamma}
(1+m)^{4-\gamma}}, \qquad 0\leq \gamma < 3
\label{dens}
\end{equation}
where
\begin{equation}
m^2=\frac{x^{2}}{a^{2}}+\frac{y^{2}}{b^{2}}+\frac{z^{2}}{c^{2}},
\qquad 0< c\leq b\leq a
\label{ellcoord}
\end{equation}
and $M$ is the total mass of the galaxy.
The gravitational potential generated by this
density law is (Chandrasekhar 1969)
\begin{equation}
\Phi({\bf r})=-\pi Gabc\int_0^{\infty} \frac
{[\psi(\infty)]-\psi(\tilde{m})]d\tau}
{\sqrt{(\tau+a^2)(\tau+b^2)(\tau+c^2)}}
\label{potchandra}
\end{equation}
with
\begin{equation}
\psi(\tilde{m})=\int_0^{\tilde{m}^2}\rho(m'^2)dm'^2,
\label{eqpsi}
\end{equation}
\begin{equation}
\psi(\infty)=\lim_{\tilde{m} \rightarrow \infty} \psi(\tilde{m}),
\end{equation}
and
\begin{equation}
\tilde{m}^2(\tau)=\frac{x^2}{a^2+\tau}+\frac{y^2}{b^2+\tau}+\frac{z^2}{c^2+\tau}.
\label{eqm}
\end{equation}
Substituting $s=\sqrt{1+\tau}$ in equation~(\ref{potchandra})
leads (for $\gamma
\neq 2$) to an integral better suited to numerical evaluation
(Merritt \& Fridman 1996, hereafter MF96):
\bea
\nonumber & \Phi(\mathbf{r})=-\frac{1}{\gamma-2} \times \\
& \times \int_0^1 \frac
{1-(3-\gamma)(\frac{\tilde{m}}{1+\tilde{m}})^{2-\gamma}+(2-\gamma)(\frac{\tilde{m}}{1+\tilde{m}})^{3-\gamma}}
{\sqrt{[1+(b^2-1)s^2][1+(c^2-1)s^2]}}ds
\eea
with
\begin{equation}
\tilde{m}^2=s^2[x^2+\frac{y^2}{1+(b^2-1)s^2}+\frac{z^2}{1+(c^2-1)s^2}].
\label{ms}
\end{equation}
For $\gamma=2$ the potential is
\bea
\Phi(\mathbf{r})= \int_0^1 \frac
{1/(1+\tilde{m})-log[(1+\tilde{m})s/\tilde{m}]ds}
{\sqrt{[1+(b^2-1)s^2][1+(c^2-1)s^2]}}+C, \\
C=\int_0^1 \frac
{logt dt}
{{\sqrt{[1+(b^2-1)t^2][1+(c^2-1)t^2]}}}.
\eea

The components of the force may be written
\begin{eqnarray}
& & \frac{\partial \Phi}{\partial x_i}\nonumber \\
&=&(3-\gamma)\frac{x_i}{a_i} \int_0^1\frac
{s^2ds}
{m_i^{\gamma}(1+m_i)^{4-\gamma}\sqrt{(a_i^2+A_1s^2)(a_i^2+A_2s^2)}}
\label{f_i}
\end{eqnarray}
where
\begin{equation}
m_i^2(s)=s^{2}\left[\frac{x^2}{a_i^2+C_1s^2}+\frac{y^2}{a_i^2+C_2s^2}+\frac{z^2}{a_i^2+C_3s^2}\right].
\label{eqmforce}
\end{equation}
Here we have used the notation
$x_1=x$, $x_2=y$, $x_3=z$, $a_1=a$, $a_2=b$,
$a_3=c$.
The constants in equations~(\ref{f_i}) and (\ref{eqmforce}) are
\begin{equation}
\begin{array}{ll}
i=1:\\
\qquad A_{1}=b^{2}-1 \qquad A_{2}=c^{2}-1\qquad\\
\qquad C_{1}=0\qquad C_{2}=b^{2}-1\qquad C_{3}=c^{2}-1\\
\\
i=2:\\
\qquad A_{1}=c^{2}-b^{2} \qquad A_{2}=1-b^{2} \\
\qquad C_{1}=1-b^{2} \qquad C_{2}=0 \qquad C_{3}=c^{2}-b^{2}\\
\\
i=3:\\
\qquad A_{1}=1-c^{2} \qquad A_{2}=b^{2}-c^{2} \\
\qquad C_{1}=1-c^{2} \qquad C_{2}=b^{2}-c^{2} \qquad C_{3}=0.\\
\\
\end{array}
\label{coeffgrav}
\end{equation}
The integrals were computed by means of the double-precision
FORTRAN routine DGAUSS of the CERNLIB library, which implements
the classic Gaussian quadrature formula.

Hereafter we adopt units in which $G=M=a=1$.
The unit of time is $T_u=a^{3/2}/\sqrt{GM}$, or
\begin{equation}
T_u=1.49 \times 10^6 \bigg{(} \frac{a}{\rm kpc}
\bigg{)}^{3/2} \bigg{(} \frac{M}{10^{11}M_{\odot}} \bigg{)}^{-1/2}{\rm yr}.
\label{units2}
\end{equation}
The unit of velocity is $V_u=\sqrt{GM/a}$, or
\begin{equation}
V_u=666.8 \sqrt{\frac{M/10^{11}M_{\odot}}{a/{\rm kpc}}} {\rm km\ } {\rm s}^{-1}.
\label{units3}
\end{equation}

\subsection{Self-consistent solutions}
\clearpage
\begin{figure*}[ht]
\includegraphics[scale=0.75]{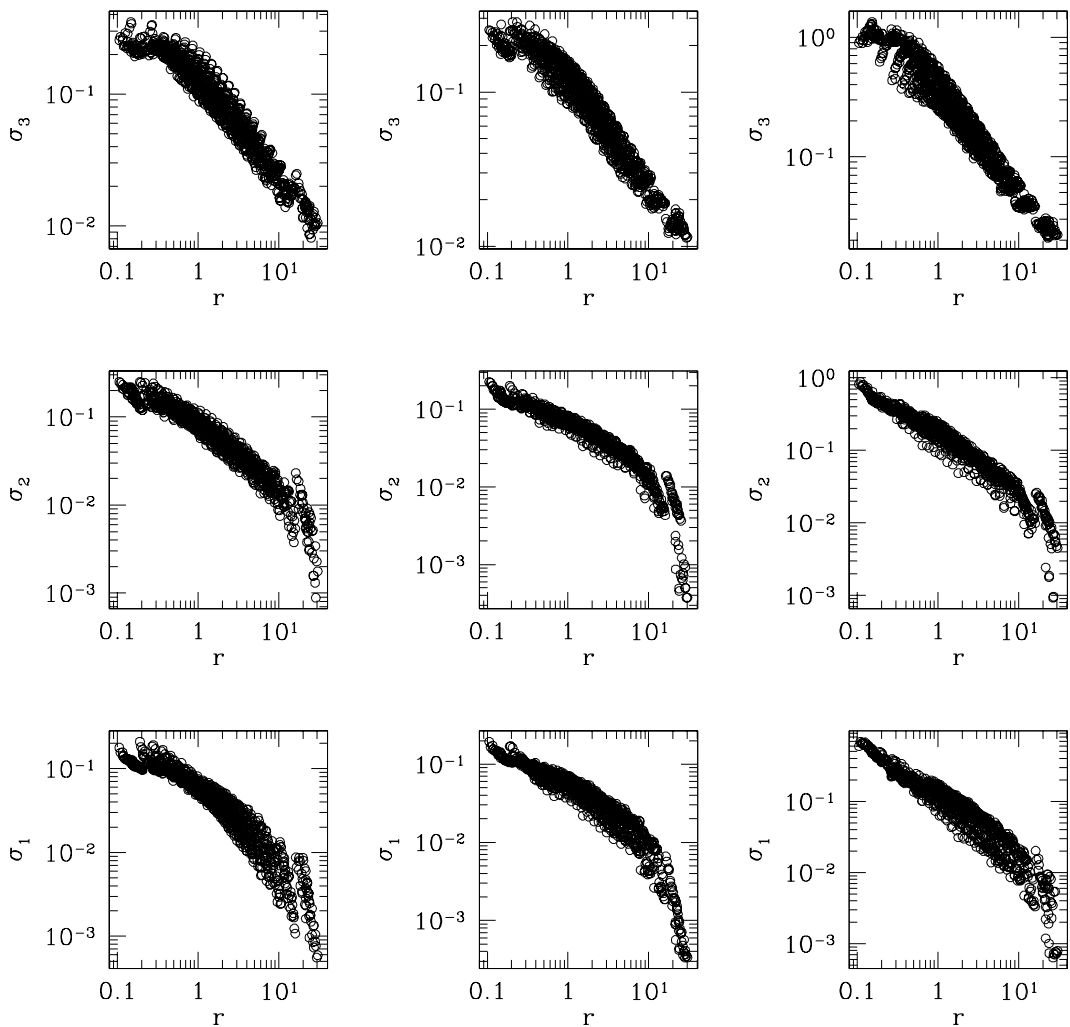}
\caption{The radial behavior of the principal components of the
velocity dispersion tensor in the three self-consistent models.
The left column refers to model 1, the central column to model 2 and 
the right column to model 3 (see Table \ref{modgal}).}
\label{disp.eps}
\end{figure*}
\clearpage
The magnitude of the dynamical friction force depends on the
details of the stellar velocity distribution.
In order to accurately follow the orbital evolution of an
ejected BH,
a self-consistent galaxy model is therefore required.

We constructed self-consistent triaxial models
via a modification of the technique described in MF96.
The densities generated by a catalogue of $M$ orbits were recorded
in a spatial grid of $N=1008$ cells; the grid
was defined as in MF96.
We then sought the linear combination of orbits,
with non-negative weights, which best 
reproduced the known mass in each cell imposing, as additional constraint, 
a zero streaming velocity.
The quantity to be minimized was
\begin{equation}
\label{chi}
\chi^2=\frac{1}{N}\sum_{l=1}^N\bigg{(}D_l-\sum_{i=1}^m C_iB_{il}\bigg{)}^2.
\end{equation}
In this expression, $C_i$ is the (unknown) number
of stars on orbit $i$ ($1\leq i \leq M$);
$B_{il}$ is the mass which the $i$th orbit
places in the $l$th cell;
and $D_l$ is the (known) mass which the model places in the $l$th cell.
The constraints $C_i \geq 0$ were also imposed, i.e.,
the number of stars on each orbit was required to be
positive.
We used the NAG quadratic-programming routine E04NCF
to carry out the minimization.

We solved the self-consistency problem for the case
$\gamma=1$, a ``weak'' density cusp, and $\gamma=2$, a ``strong'' cusp.
In the $\gamma=1$ case, we built two models with different axis ratios,
as reported in Table \ref{modgal}.
All models have a ``triaxiality index'' $T=0.5$,
where $T\equiv (a^2-b^2)/(a^2-c^2)$; i.e. they are
``maximally triaxial''.
The weak-cusp model is similar in structure to bright
elliptical galaxies and bulges, those having
absolute visual magnitudes
brighter than $M_V\approx -21$, while the strong-cusp
model is similar to low-luminosity spheroids like
that of the Milky Way and M32, which have higher-density
nuclei.
Table~\ref{modgal} also gives, as useful reference time,
the crossing time $T_{1/2}$,
defined as the period of the circular orbit at the radius
containing one-half of the total galaxy mass,
in the ``equivalent'' spherical model; the latter
is defined as the spherical model with scale length
given by $(abc)^{1/3}$ and the same total mass.

In order to construct the velocity dispersion tensor, 
which is needed in the computation of the dynamical friction
force, we also stored the velocities of
the orbits as they passed through the cells.
The velocity dispersion tensor is (Merritt 1980)
\begin{equation}
\sigma^2_{jk,l}=
\frac
{\displaystyle\sum_{i=1}^N C_i B_{il}<V^2_{jk}>_{il}}
{\sum_{i=1}^N C_i B_{il}}
\label{eqsigma}
\end{equation}
where 
\begin{equation}
<V^2_{jk}>_{il}=<V_jV_k>_{il}-<V_j>_{il}<V_k>_{il};
\label{vjk}
\end{equation}
here $j,k$ refer to the coordinate axes,
$i$ is the orbit number, and $l$ is the cell number.
The principal components of the velocity dispersion tensor,
$\sigma_k^2, k=1,2,3$, in each cell could then be computed via
standard techniques.

The high-energy cutoff in the orbital sample together with the unavoidable limitation
in abundance of the orbital catalog implies that the outermost region of the spatial
grid is not well populated, so we excluded
from our evaluation of $\sigma^2_{jk,l}$ the 48 cells of the outermost spatial shell.

Figure~\ref{disp.eps} shows the radial behavior of the principal
components of the velocity dispersion tensor in each of the three 
self-consistent models.
\clearpage
\begin{deluxetable}{cllll}
\tablewidth{0pt}
\tablehead{
\colhead{Model} & \colhead{$\gamma$} & \colhead{$b/a$} & \colhead{$c/a$}
& \colhead{$T_{1/2}$} }
\startdata
1 & 1 & 0.7906 & 0.5 & 21.0\\
2 & 1 & 0.8631 & 0.7 & 25.9\\
3 & 2 & 0.8631 & 0.7 & 6.91\\
\enddata
\label{modgal}
\end{deluxetable}

\begin{deluxetable}{cccccc}
\tablewidth{0pt}
\tablehead{
\colhead{ID} & \colhead{$\gamma$} & \colhead{$r_{c}/r_{BH}$\tablenotemark{a}}
& \colhead{Core\tablenotemark{b}} & \colhead{$v_{e}$\tablenotemark{c}} &
\colhead{$v_{e,eff}/v_{e}$\tablenotemark{d}}
}
\startdata
1a & 1 & 0   & I   & 1.625 & 1.018 $-$ 1.044\\
1d & 1 & 0.2 & II & 1.620 & 1.008 $-$ 1.036\\
1e & 1 & 2   & II & 1.605 & 1.001 $-$ 1.023\\
1f & 1 & 0.2 & III  & 1.620 & 1.008 $-$ 1.030\\
1g & 1 & 2   & III & 1.605 & 1.001 $-$ 1.017\\
2a & 1 & 0   & I   & 1.530 & 1.030 $-$ 1.047\\
2d & 1 & 0.2 & II & 1.530 & 1.012 $-$ 1.028\\
2e & 1 & 2   & II & 1.515 & 1.004 $-$ 1.019\\
2f & 1 & 0.2 & IV  & 1.530 & 1.009 $-$ 1.024\\
2g & 1 & 2   & IV  & 1.515 & 1.004 $-$ 1.013\\
3b & 2 & 0.2 & III  & 4.66  & 1.73  $-$ 2.15\\
3c & 2 & 2   & III  & 4.04  & 1.89  $-$ 2.24\\
3d & 2 & 0.2 & II & 4.66  & 2.20  $-$ 3.20\\
3e & 2 & 2   & II & 4.04  & 1.11  $-$ 1.35\\
3f & 2 & 0.2 & III  & 4.66  & 1.08  $-$ 1.21\\
3g & 2 & 2   & III & 4.04  & 1.00  $-$ 1.05\\
\enddata
\tablenotetext{a}{Core radius in units of $r_{BH}$ (Eq.~\ref{eq:rbh}).}
\tablenotetext{b}{Scheme for treatment of central forces (see \S2.2).}
\tablenotetext{c}{Central escape velocity neglecting dynamical friction.}
\tablenotetext{d}{Central escape velocity including dynamical friction.}
\label{modincen}
\end{deluxetable}
\clearpage
\subsection{The dynamical friction force}

The classical Chandrasekhar (1943) formula for the dynamical friction
force, which assumes a homogeneous and isotropic medium,
has been extended by Pesce, Capuzzo-Dolcetta \& Vietri (1992)
to the triaxial case, in partial analogy with the Binney's treatment of the
axysimmetric case (Binney 1977).
The result is
\begin{equation}
\mathbf{f}_{df,t}=-\Gamma_{1} \mathbf{\widehat{e}}_1-\Gamma_{2}
\mathbf{\widehat{e}}_2-\Gamma_{3} \mathbf{\widehat{e}}_3
\label{adf}
\end{equation}
\noindent
where $\mathbf{\widehat{e}}_k$ ($k=1,2,3$) are
the principal components of the velocity dispersion tensor and
$\Gamma_k(\mathbf{r},\mathbf{v},v_{k})=\gamma_k(\mathbf{r},\mathbf{v})v_k$,
with $v_k$ the component of the BH's velocity along
the $\mathbf{\widehat{e}}_k$ axis.
The functions $\gamma_i(\mathbf{r},\mathbf{v})$ are given by

\bea
\label{gammai}
\displaystyle
\nonumber \gamma_i(\mathbf{r},\mathbf{v})= &2\sqrt{2\pi}\rho (\mathbf{r})G^{2}\ln \Lambda (m +
\mh) \sigma _{3}^{-3} \times \\
& \int_{0}^{\infty} \frac{e^{\Sigma_{k=1}^{3} -\frac{v_{k}/2\sigma
_{k}^{2}}{\epsilon _{k}^{2}+u}}}{(\epsilon _{i}^{2}+u)\sqrt{\Sigma_k (\epsilon
_{k}^{2}+u)}}du.
\eea
\noindent

Here, $m_*$ is the mass of a field star,
ln$\Lambda$ is the usual Coulomb logarithm, $\mh$
is the BH mass, $G$ is the gravitational
constant, $\rho({\bf r})$ is the galaxy mass density,
and $\epsilon_k$ is the ratio between $\sigma_k$
and the largest eigenvalue, defined to be $\sigma_3$.
\clearpage
\begin{figure}
\includegraphics[scale=0.7]{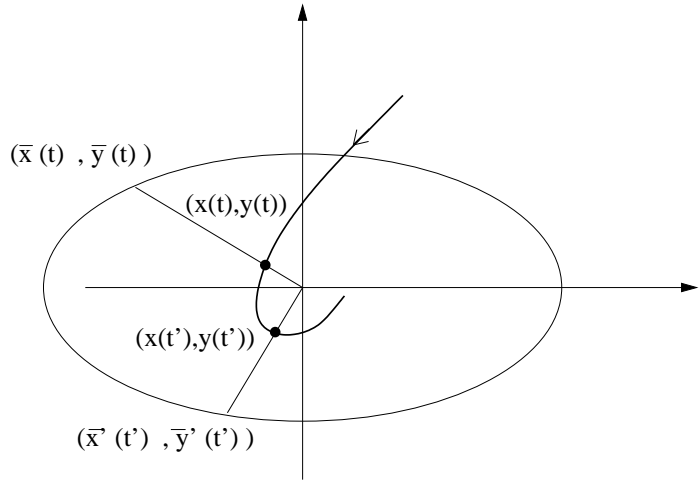}
\caption{A two-dimensional illustration of the definition of
$\bar{x}$ and $\bar{y}$; the solid curve is the orbit of the BH.
The ellipse has equation $m^2=\bar{m}^2$.}
\label{tr2sf}
\end{figure}
\clearpage
The motion of the test object (the black hole) can be strongly affected by the
values of the conservative and dynamical friction forces very near
the center.
Both components of the force would be influenced by the motion of the
test particle; for instance, displacing a BH causes the central
density to drop, an effect that was ignored in the construction
of the models.
In the case of the dynamical friction force,
the limited resolution of our triaxial models is a problem;
the central cells have a
size of $0.28$ and $0.05$ for $\gamma =1$ and $\gamma =2$, respectively.
We adopted the following scheme to deal with these issues.

The standard Chandasekhar (1943) formula in the so called local approximation 
evaluates the dynamical
friction force using the values of the stellar density
and velocity at the position of the test particle.
In the case of a particle that is at the center of a
galaxy with a steep central density profile,
the local approximation clearly yields an overestimate of the actual
dynamical friction force, because it does not weight properly 
the contributions of particles in zones surrouding the
center, where the density is much smaller.
To overcome this problem, Capuzzo-Dolcetta \& Vicari (2006)
proposed, for the case of a centrally-located test particle of mass $\mh$,
the use of a numerical evaluation of the dynamical friction 
integral in its complete form.
The fact that the test particle is at the center is exploited by
identifying the distance to the scatterer ($r$) with
the impact parameter $b$, yielding
\bea
\nonumber
{\mathbf f}_{df}= -4\pi \frac{m}{\mh+m} \times \nonumber \\
\int_{b_{min}}^{b_{max}}\int_{{\mathbf v}}
f(b,\mathbf{v})
\frac{({\mathbf v}_{\bullet}-{\mathbf v})|{\mathbf v}-{\mathbf v}_{\bullet}|}
{1+\frac{b^2|{\mathbf v}-{\mathbf v}_{\bullet}|^4}{G^2(\mh+m)^2}}
d^3\mathbf{v} b db\equiv & \nonumber \\
-G(v_\bullet)\frac{{\mathbf v}_{\bullet}}{v_{\bullet}}&,
\label{dfcorrect}
\eea
\noindent
where $f(\mathbf {r},\mathbf{v})$ is the distribution function
of the background particles (of mass $m$) that provide the
frictional force.
The integral in (\ref{dfcorrect}) is
numerically performed assuming $b_{min}=0$ and, for $b_{max}$, 
a value large enough to guarantee convergence. 

After fitting with a suitable analytic expression the numerical
$G(v_\bullet)$, we obtain an estimate of the dynamical friction
force in the inner ($m \leq \bar{m} \equiv 0.05$) region of the galaxy
by means of the interpolating formula
\begin{equation}
{\mathbf f}_{df}=-p(m) G(v_\bullet) \frac{{\mathbf
v}_{\bullet}}{v_{\bullet}} + [1-p(m)]
{\mathbf f}_{df,t}(\bar{x},\bar{y},\bar{z}),
\label{adf3}
\end{equation}
where $p(m)$ is a regular weighting function, satisfying $p(0)=1$ and
$p(\bar{m})=0$. The expression \ref{adf3} connects the central dynamical friction force,
equation~(\ref{dfcorrect}),
with its values at $m > \bar{m}$, obtained via equation~(\ref{adf}).
Here, $\bar{x}(t),\bar{y}(t)$ and $\bar{z}(t)$ are the coordinates
of the intersection point between the radial direction
through the position of the test particle at time $t$ and the $m^2=\bar{m}^2$
ellipsoidal surface (Figure~\ref{tr2sf}). In the following, we assumed for 
simplicity a linear dependence of $p(m)$ on $m$.

So far, we have ignored the influence of the BH and its motion
on the distribution of mass in the galaxy.
In reality, sudden removal of the BH from its central location
will cause the stellar density to drop.
The ejected BH will carry with it stars initially contained
within a region $r<r_{eff}$ such that the orbital velocity around the BH
is equal to $V_{kick}$, or
\begin{equation}
r_{eff} \approx \frac{G\mh} {V_{kick}^2} \approx
r_{BH}\left(\frac{V_{kick}}{sigma}\right)^{-2},
\end{equation}
with
\begin{equation}
r_{BH}\equiv G\mh/\sigma^2
\label{eq:rbh}
\end{equation}
the BH gravitational influence radius.
Removal of the BH also reduces the gravitational force that
binds the stars at $r\gap r_{BH}$, causing them to move outward.
Both effects result in a lowering of the density at $r\lap r_{eff}$.

A self-consistent treatment of this expansion would require
$N$-body techniques.
Here, we account approximately for the expansion by considering 
the dependence of our results on different assumptions about 
the inner form of the stellar density profile.
We introduce a ``core'' radius $r_c$, parametrized in terms of $r_{BH}$ 
as $r_c=\alpha r_{BH}$, $\alpha=(0.2,2)$.
In computing $r_{BH}$, we set in Eq. \ref{eq:rbh} as
$\sigma$ the value of the velocity dispersion in the innermost
model cell.
\clearpage
\begin{figure}
\resizebox{\hsize}{!}
{\includegraphics[scale=0.7]{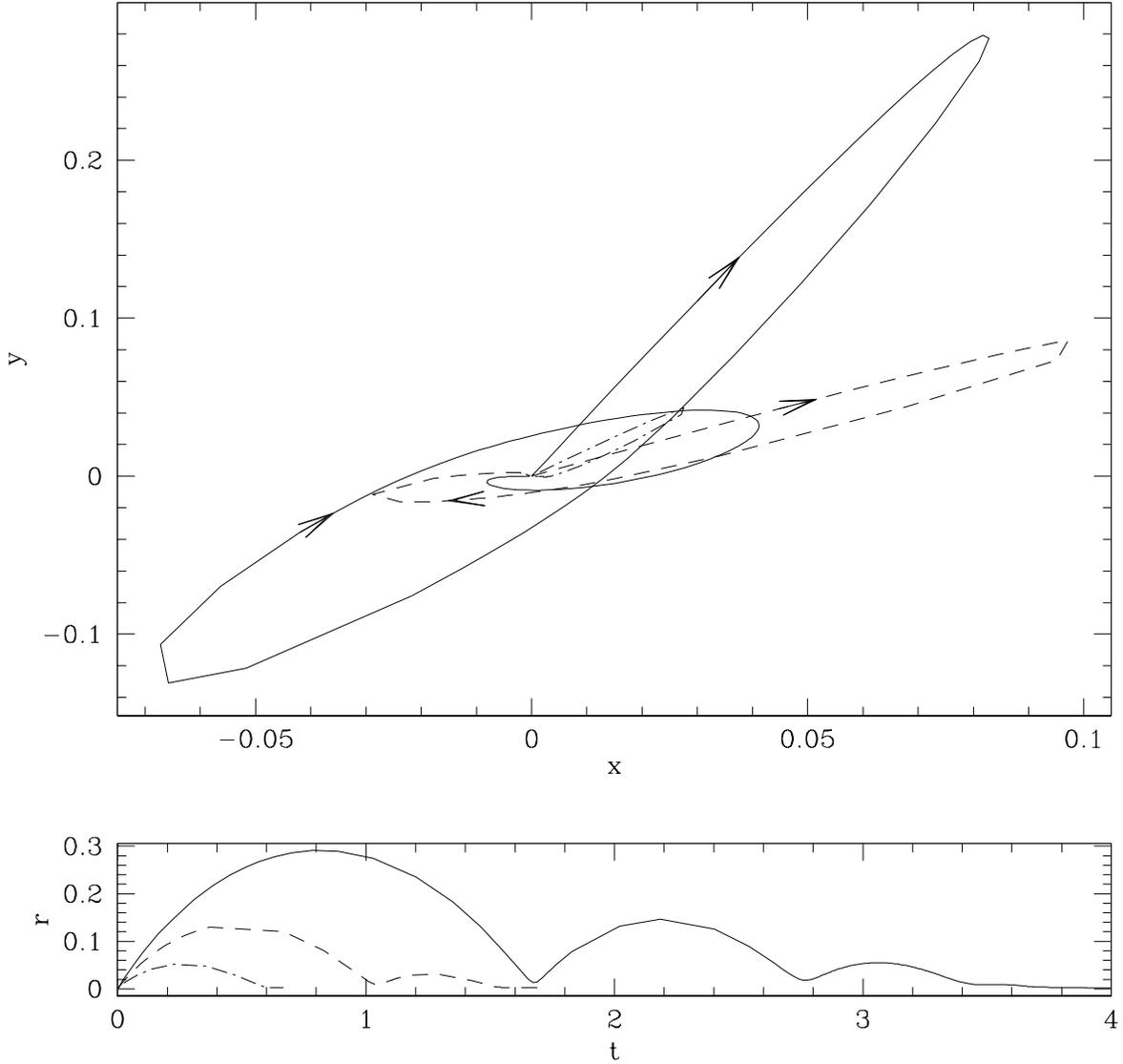}}
\caption{Evolution of three planar ($x=0$) orbits in model 2a
($\gamma=1$), for $V/v_e=(0.4, 0.5, 0.6)$; for display convenience, we chose
orbits with a different initial $\theta$.
Arrows indicate the direction of the motion for the two more energetic orbits.
The lower panel shows the distance of the BH
from the center as function of time.}
\label{fig:ex2}
\end{figure}

\begin{figure}
\resizebox{\hsize}{!}
{\includegraphics{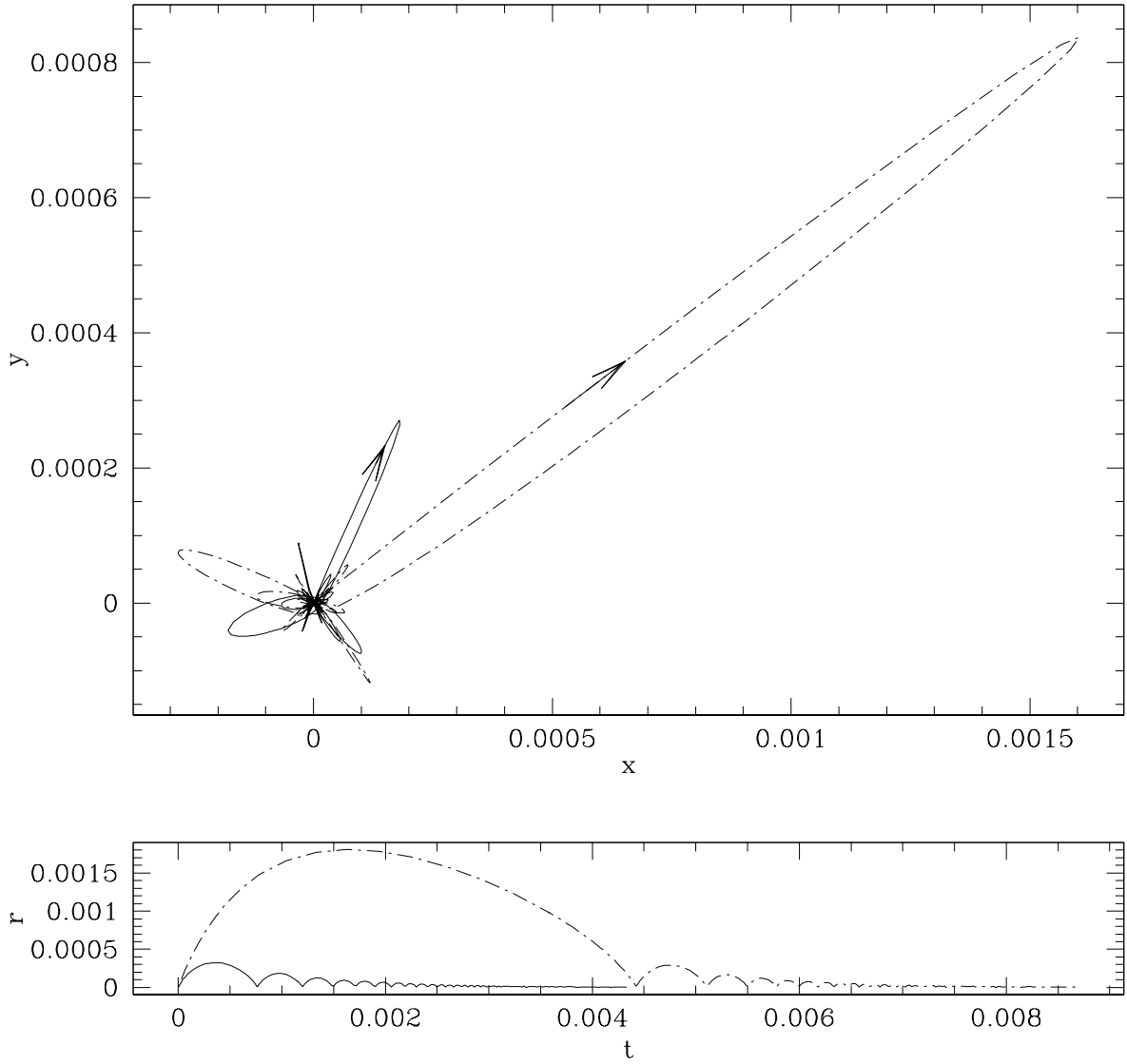}}
\caption{As in Figure~\ref{fig:ex2}, for two planar orbits in
model 3b ($\gamma=2$), with $V/v_e=(0.8, 0.9)$.}
\label{fig:ex3}
\end{figure}

\begin{figure*}
\includegraphics[width=\textwidth]{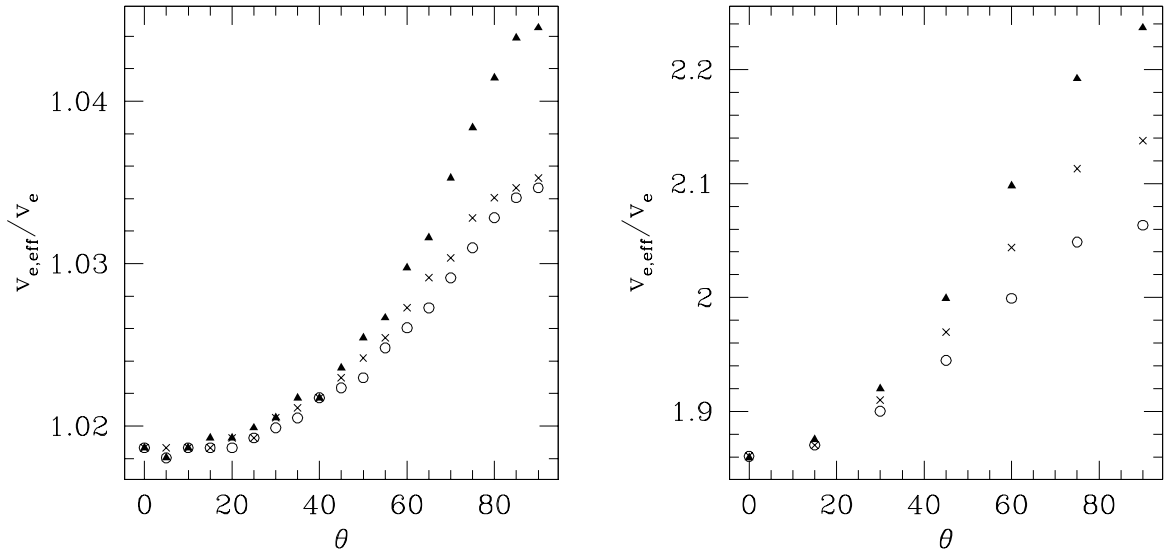}
\caption{Effective escape velocity as function of launching angle
$\theta$, for model 1a (left panel) and 3c (right panel).
Empty circles refer to $\phi=90^\circ$, crosses to $\phi=45^\circ$,
and black triangles to $\phi=0^\circ$.}
\label{fig:veff}
\end{figure*}
\clearpage

Note that $r_{BH}$ is \lq under\rq~ the model resolution in all the cases treated,
being $r_{BH}/L=0.16$ and $r_{BH}/L=0.025$ in the cases of $\gamma=1$ and $\gamma=2$,
respectively ($L$ is the innermost cell size, different in the two cases).

Based on these considerations, we adopted four different schemes
for treating the central forces (conservative + frictional),
which we denote in Table~\ref{modincen} via the labels I-IV.
These schemes are defined as follows.

\begin{enumerate}
\item[I]
No modification to the density profile, i.e. $r_c=0$.
Conservative forces were computed as in equation~(\ref{f_i})
(which is possible just when the central forces are finite, i.e.
$\gamma=1$) and dynamical friction forces as in equation~(\ref{adf3}),
i.e. by matching the triaxial expression to the central
expression.
\item[II]
Conservative forces within the core were set to zero.
Frictional forces in this region were computed via
equation~(\ref{adf}), assuming for the coefficients
$\Gamma_k$ their values at the boundary $r=r_c$, as averaged over
angles, and with the correct velocity dependence, i.e.
$\Gamma_k=<\Gamma_k(r_c, {\bf v}, v_k)>_{\theta,\phi}$.
\item[III]
Density was set to zero within the core,
thus conservative forces zero therein,
and frictional forces in the core were
evaluated setting $b_{min}=r_c$ and $p(m)=1$ in 
equation~(\ref{adf3}).
\item[IV]
The density was given a core of radius $r_c$ and
the potential within the core was assumed harmonic
(linear forces), matching the external triaxial potential at the core 
boundary. Frictional forces were computed as in I.
\end{enumerate}

In what follows, we assumed a BH mass of $\mh=10^{-3}$ in units of
the galaxy mass, close to the mean ratio observed in real galaxies
(Merritt \& Ferrarese 2001; Marconi \& Hunt 2003).
The Coulomb logarithm was set to $\ln \Lambda=6.6$ (Spinnato et al. 2003).
Dynamical friction times scale linearly with $(\mh\ln \Lambda)^{-1}$.

\section{Results}
We studied the orbital evolution of the massive particle (BH)
after being ejected from the center of each of the galaxy models in Table \ref{modincen}
with different kick velocities $V$ and angles ($\theta,\phi$),
where $V_x=V\sin\theta\cos\phi$, $V_y=V\sin\theta\sin\phi$, $V_z=V\cos\theta$.
The kick velocity $V$ was assigned a value in the range
$0.2v_e\leq V \leq v_{e,eff}$, where
$v_e$ is the escape velocity from the origin in the
absence of frictional forces, and $v_{e,eff}$ is the actual
velocity required for escape; $v_{e,eff}>v_e$ since dynamical
friction acts to reduce the particle's kinetic energy.
Table \ref{modincen} gives $v_e$ and $v_{e,eff}/v_e$
(the latter expressed as a range, since $v_{e,eff}$ depends on the
launching angle) for each of the models.
The direction of the kick was given one of 43 values
by choosing $\theta$ and $\phi$ from the discrete set ($0^{\circ},
15^{\circ}, 30{^\circ}, 45{^\circ}, 60{^\circ}, 75{^\circ}, 90{^\circ}$).
We defined the decay time $T_{df}$ as the time
when the BH orbital energy had dropped to 1\% its initial value
and its
residual time variation was negligible ($|\dot E/E_0| \leq 10^{-4}$).

Figures~\ref{fig:ex2} and \ref{fig:ex3} show the evolution of
a representative set of orbits.
Figure~\ref{fig:ex2} shows that for moderate kicks ($V\lap 0.4 v_e$)
the BH executes only one ``bounce'' before dynamical friction
brings it to a halt at the center.
Figure~\ref{fig:ex3} illustrates some higher energy orbits ($V/v_e=0.8,0.9$)
in a strong-cusp model ($\gamma=2$).
In spite of the strong frictional forces near the center of this model,
ejection with a sufficiently large $V$ allows the BH to execute several oscillations
before coming to rest.

The dependence of the effective escape velocity $v_{e,eff}$
on launching angle for two galaxy models is illustrated in
Figure~\ref{fig:veff}.
The increase in $v_{e,eff}$ relative to $v_e$ is most striking
for trajectories aligned with the $x-$ (long) axis, and
in models with high central densities ($\gamma=2$);
$v_{e,eff}/v_e$ can be as large as $\sim 3$
(Table \ref{modincen}).
However, Table \ref{modincen} also shows that $v_{e,eff}/v_e$ for these
high density models can depend substantially on how
the central forces are treated.
For this reason, we consider the largest values of $v_{e,eff}/v_e$
in Table \ref{modincen} to be provisional, until they can be verified with
detailed $N$-body simulations.
\clearpage
\begin{figure*}
\includegraphics[totalheight=0.35\textheight,viewport= 0 0 592 280,clip]{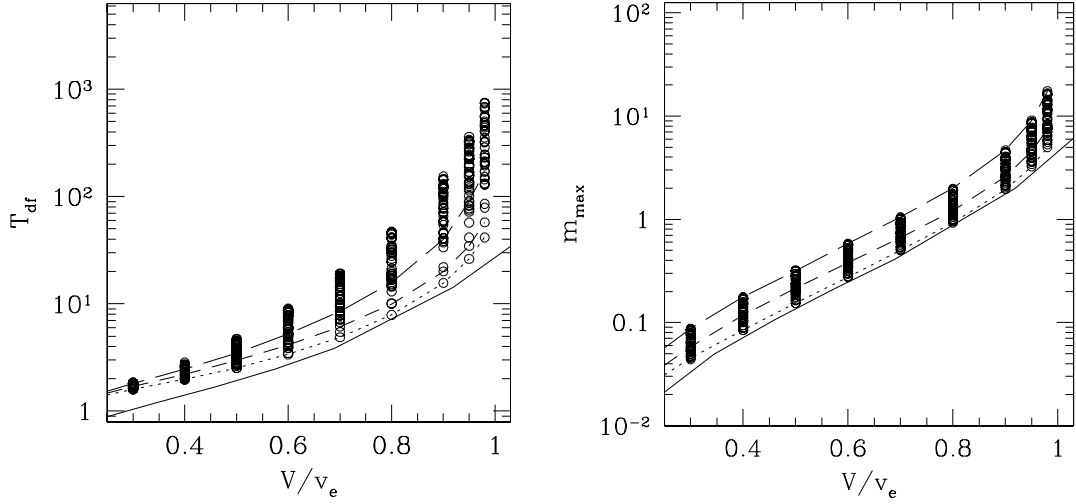}
\caption{{\it Left panel}: Decay time $T_{df}$ of the BH
 as function of the kick velocity $V$ and for several launching angles
(empty circles). In both panels, the dotted, short-dashed and long-dashed lines
 refer to kicks along the x-axis, y-axis and z-axis, respectively. The solid
lines are for the ``equivalent'' spherical model, with scale length $(abc)^{1/3}$.
{\it Right panel}: the maximum displacement of the BH.
In both panels, dotted lines are for the ``equivalent'' spherical model,
with scale length $(abc)^{1/3}$.
All results in this figure refer to model $1a$.}
\label{metvsv1}
\end{figure*}

\begin{figure*}
\includegraphics[totalheight=0.35\textheight,viewport= 0 0 592 280,clip]{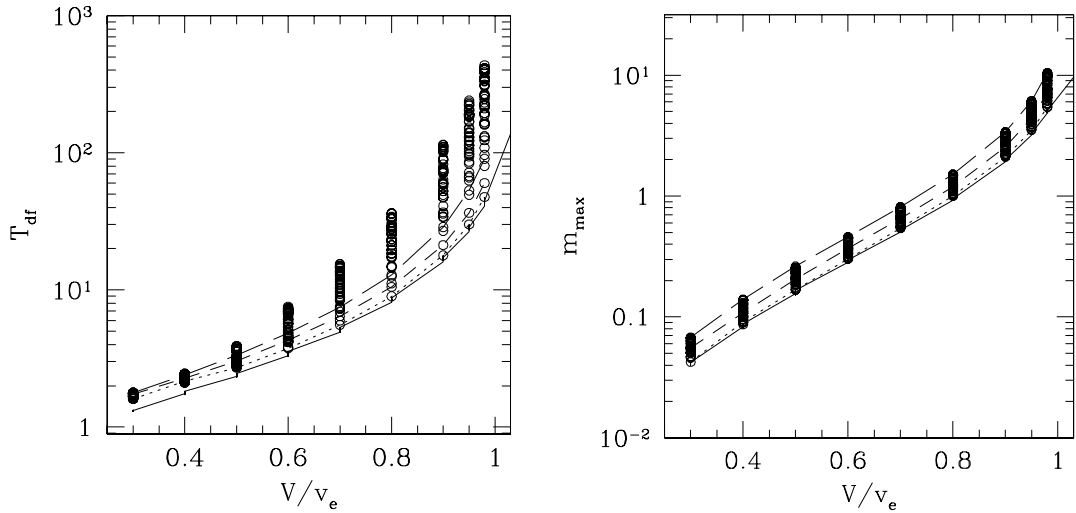}
\caption{Like Figure~\ref{metvsv1}, but for model $2a$.}
\label{metvsv2}
\end{figure*}

\begin{figure*}
\includegraphics[totalheight=0.35\textheight,viewport= 0 0 592 280,clip]{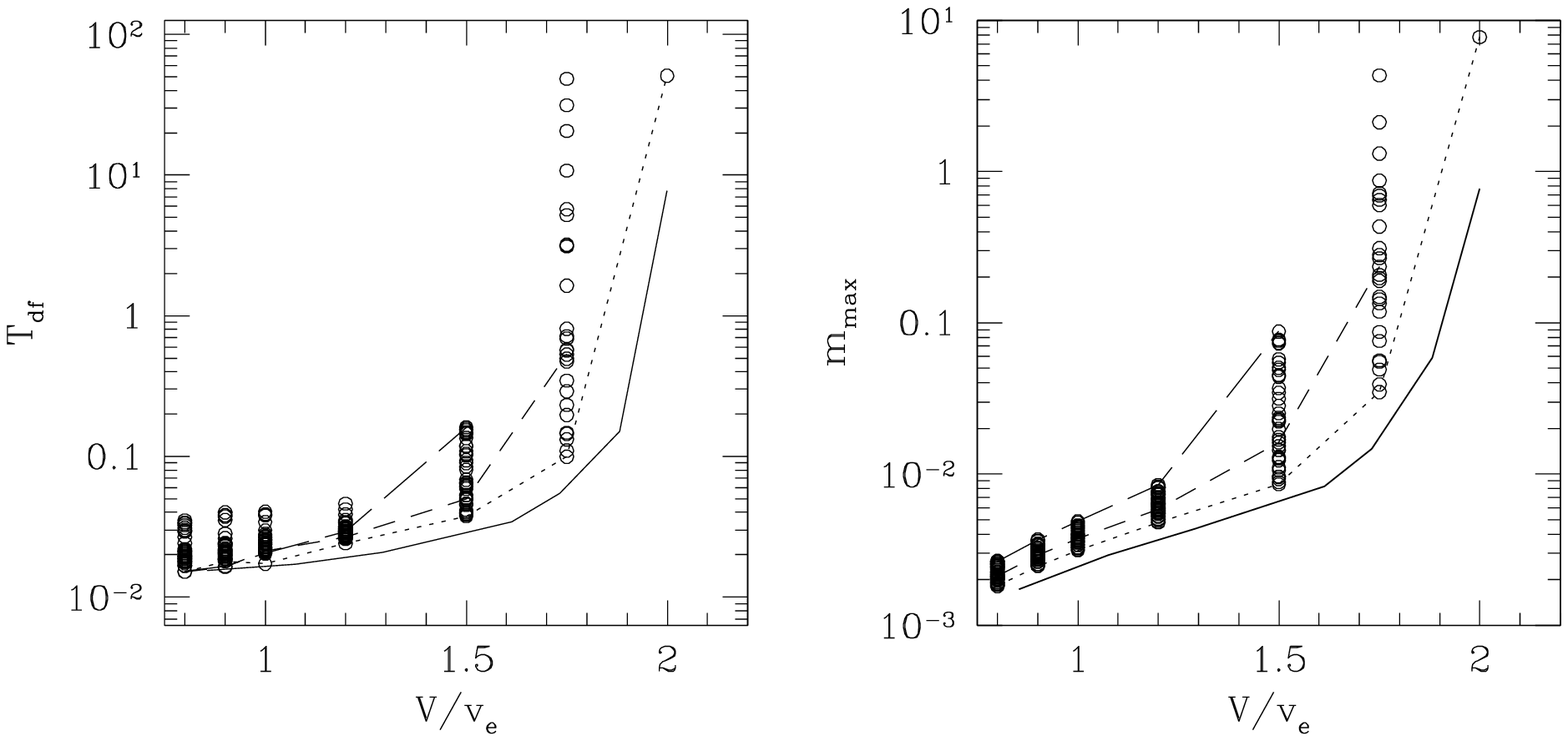}
\caption{Like Figure~\ref{metvsv1}, but for model $3b$.}
\label{metvsv3b}
\end{figure*}

\begin{figure*}
\includegraphics[totalheight=0.35\textheight,viewport= 0 0 592 280,clip]{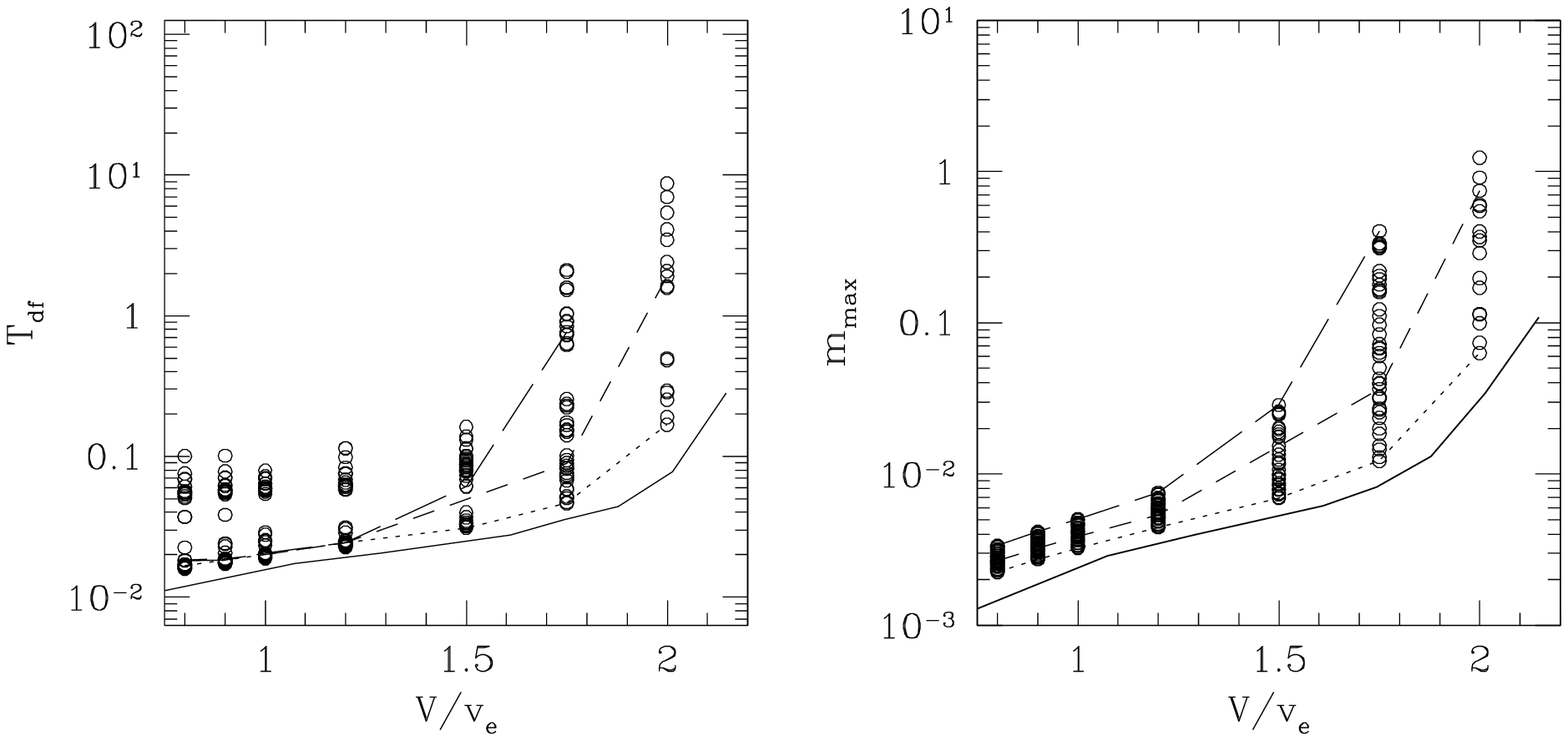}
\caption{Like Figure~\ref{metvsv3b}, but for model $3c$}
\label{metvsv3c}
\end{figure*}

\begin{figure*}
\includegraphics[width=\textwidth]{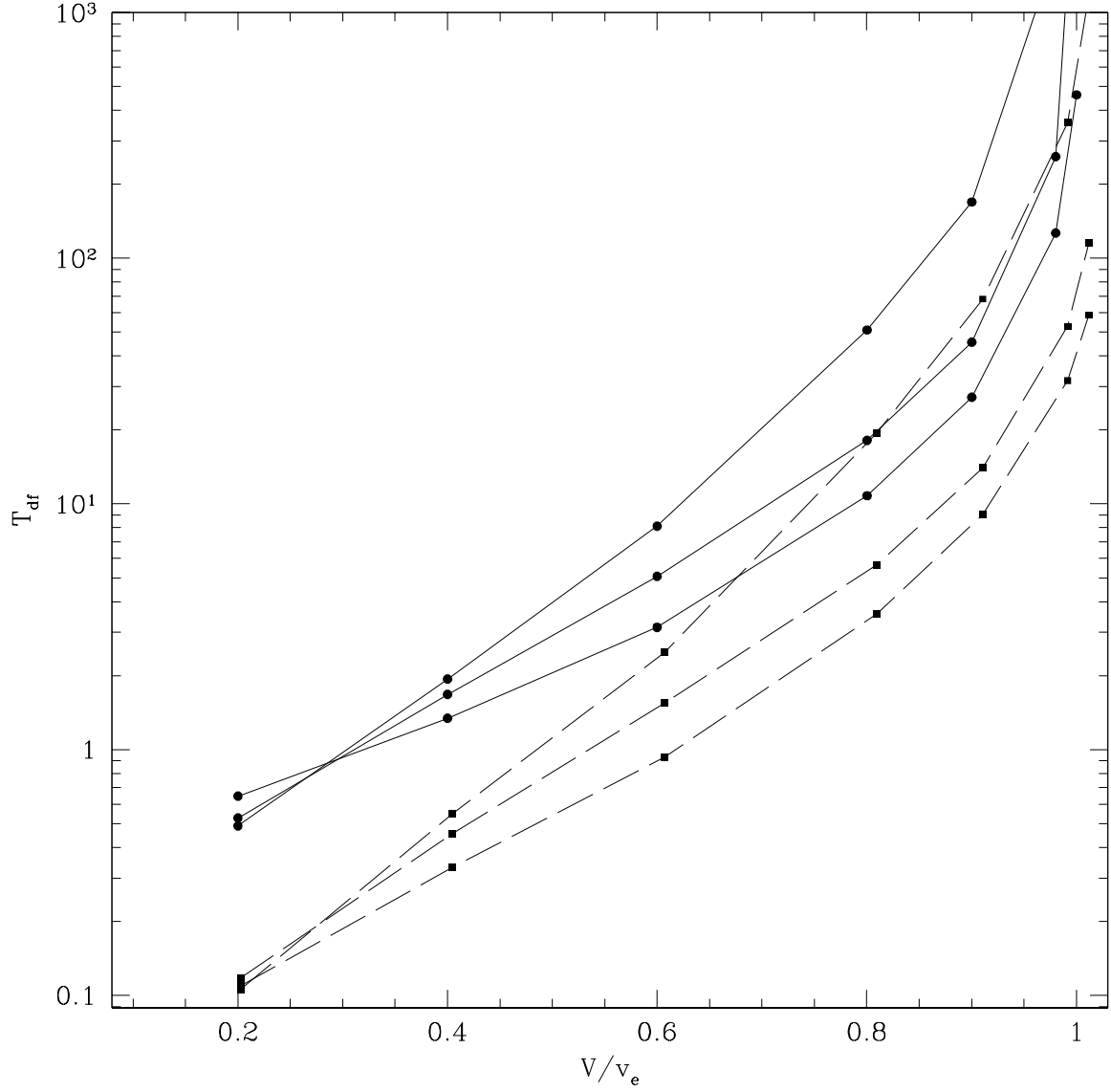}
\caption{Orbital decay times vs the kick velocity for model 1f (dashed line)
and 1g (solid line) for the BH launched along the x (quickest decay), y and z
(slowest decay) axis.}
\label{corecomp}
\end{figure*}

\begin{figure*}
\resizebox{\hsize}{!}
{\includegraphics[scale=1]{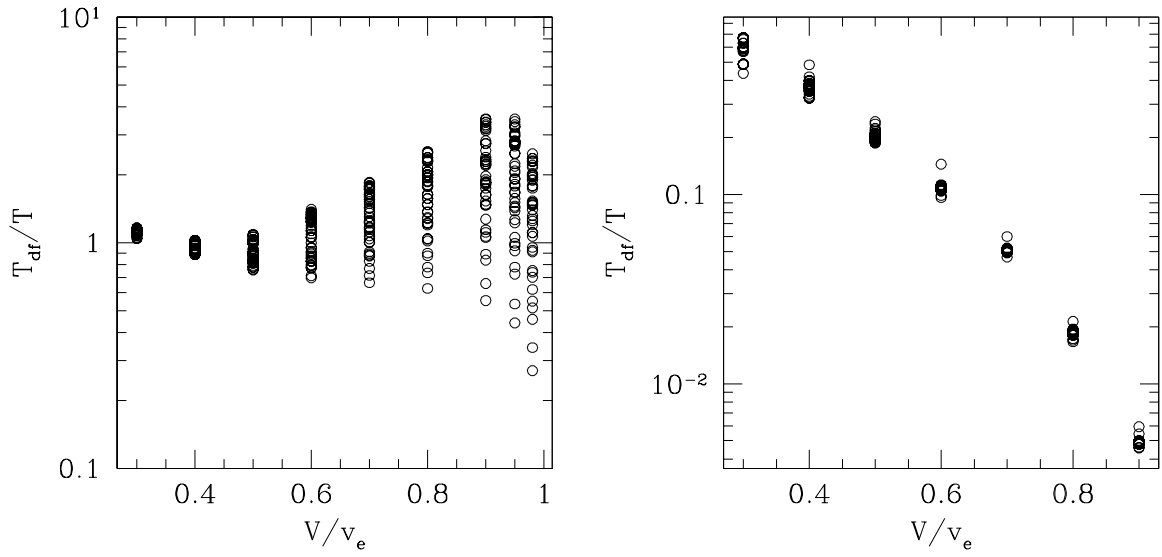}}
\caption{Orbital decay time in units of the period $T$ of
the unperturbed $x$-axial orbit of the same energy,
for models 2a (left panel) and 3b (right panel).}
\label{tdfvsv}
\end{figure*}

\begin{figure*}
\includegraphics[width=\textwidth]{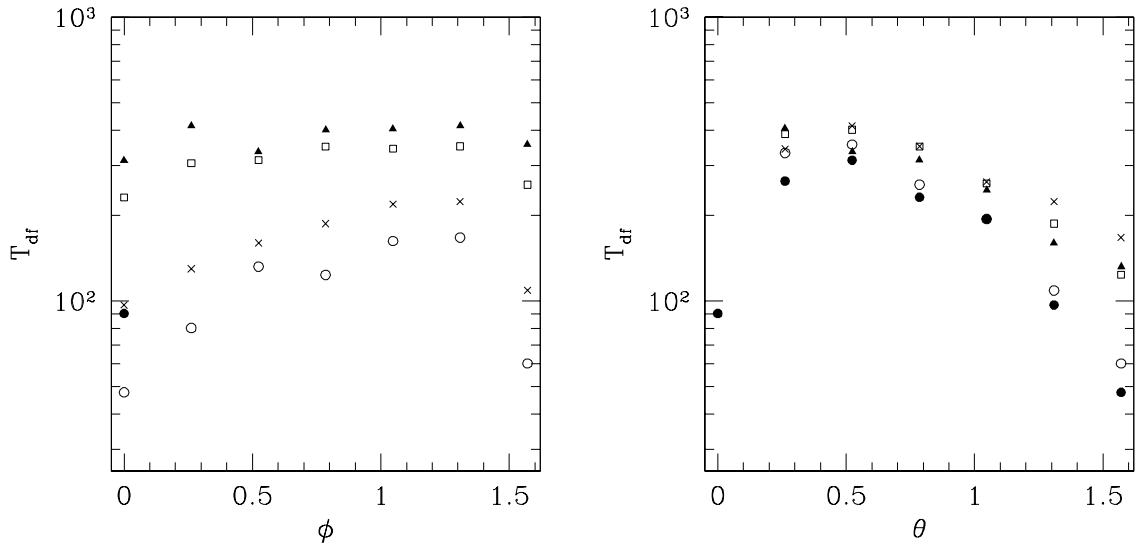}
\caption{Orbital decay times in model 1a as a function of launching angle $\phi$
(left panel) and $\theta$ (right panel), for $v_0/v_e=0.8$.
In each panel, symbols of the same type represent fixed values of the
other angle, according to the scheme: black circles = 0$^\circ$;
black triangles = 30$^\circ$;
squares = 45$^\circ$; crosses = 75$^\circ$; empty circles=90$^\circ$.}
\label{ang}
\end{figure*}
\clearpage
Figures~\ref{metvsv1}-\ref{metvsv3c} show the dependence of the
decay time, $T_{df}$, and of the apocentric ``ellipsoidal''
distance, $m_{max}$, on the initial velocity in four of the models,
for the entire set of launching angles.
For the sake of comparison, the values
of $T_{df}$ and $m_{max}$ for the ``equivalent'' spherical model,
characterized by the density law (\ref{dens}) with $r$ instead of $m$
and length scale equal to $(abc)^{1/3}$, are plotted as dotted lines.
Kicks along the symmetry axes result in the shortest
return times: the corresponding orbits are exactly radial and pass
precisely through the dense center on each return, as in
a spherical model.
%
Return times are longer than in a spherical model, by as much as an
order of magnitude in the $\gamma=1$ models, and even larger factors in the
$\gamma=2$ models.
In addition, Fig. \ref{corecomp} compares the decay times
in two models (1f and 1g) with the same treatment of the dynamical friction term
and same density profile but a different core size, a factor ten larger in mod. 1g.
Note the shorter decay times in the case of the smaller, higher density, core, due
also to the smaller elongation of the radial oscillations at a given $V/v_e$ caused by
the higher potential well.

Figures 6-10 also show that oscillations along the major axis are quenched
more rapidly than along the other axes, as demonstrated previously by
Capuzzo-Dolcetta \& Vicari (2005).
Even though orbits launched close to the $z$- (short-) axis reach the greatest
values of $m_{max}$ at every $v/v_e$, their decay times are not
necessarily the longest.
This is because the dynamical friction force transforms the
near-radial orbit into a box orbit which has a thin waist,
thinner than that of orbits having the same initial energy but
launching angles $\theta \neq 0$; the dynamical friction force
is therefore greater because of the closer approach of the
BH to the high-density regions.
This is not a trivial result, because the galactic models treated in this
paper are cuspy, and it was not {\it a priori} obvious that
the large-scale behavior of the density would exert the dominant influence
on the motion.
At first sight it is surprising that the oscillations in the
equivalent spherical models are always shorter than in the triaxial case.
since in absence of dynamical friction,
the radial oscillation in the spherical case reaches a maximum
extension which is intermediate between the displacements
along the $x$ and $z$ axes in the trixial case.
However, the return time is always shortest in the spherical geometry
since the effect of dynamical friction in the triaxial case is
progressively greater from $x$ to $z$,
making the $x$ oscillation the shortest,
but still larger than in the spherical case.

Decay times are often of the same order, or less than, the period of
an unperturbed radial orbit in the spherical geometry,
particularly in the models with $\gamma=2$.
This is illustrated in Figure~\ref{tdfvsv} for two models.
Thus, in many cases, a kicked BH executes only $\sim$ one
or fewer complete radial oscillations before coming to rest.

Figure~\ref{ang} illustrates the dependence of $T_{df}$ on the launching
angle.
Decay times peak for intermediate angles, as was already apparent
from Figures ~\ref{metvsv1}-\ref{metvsv3c}.

\section{Consequences for Black Hole Displacements}

While triaxiality has the effect of lengthening the mean
return time of a kicked BH to the center, compared
with the time in a spherical galaxy (Figs.~\ref{metvsv1}-\ref{metvsv3c}),
infall times are still typically short, of order a galaxy
crossing time, unless the kick velocity is close to
the escape velocity.
Another way to state this result is to say that
there is a narrow range of kick velocities such that the
BH spends a long time away from the center without being
fully ejected.

Here we estimate the probability that a kicked BH
will be found at an appreciable distance from the center
in a randomly-chosen galaxy.
Since the distribution of kick velocities is poorly known,
we will present results as a function of $V/v_e$.
The other parameter that determines the likelihood of finding
a displaced BH is the time $\tau$ since the galaxy
experienced its last merger (which we equate with the elapsed
time since the kick).
This time is also poorly know for any galaxy.
We therefore assume that $\tau$ follows a Poisson distribution,
\begin{equation}
p(\tau)d\tau = \frac{e^{-\tau/t_{\rm merge}}} {t_{\rm merge}} d\tau,
\label{eq:poft}
\end{equation}
with $t_{\rm merge}$ the mean time between mergers.

Let $P_V(r;\Delta t)dr$ be the probability of finding a kicked BH
a distance $r$ to $r+dr$ from the center of the galaxy at a time
$\Delta t$ after the kick.
In the case of a triaxial galaxy, we define $P_V$ as an
average over the two launch angles $(\theta,\phi)$.
Clearly for $\Delta t \ge T_{df,V}$, this distribution is a delta
function at the origin, where $T_{df,V}$ is defined to be
the maximum return time for kicks of magnitude $V$.
In a randomly-chosen galaxy, the distribution of displacements for kicks
of magnitude $V$ is
\begin{equation}
N_V(r) = \int_0^\infty p(\tau)P_V(r;\tau) d\tau.
\label{eq:nofr}
\end{equation}

We simplify this expression by assuming that
$T_{df}$ is short compared with $t_{merge}$,
allowing us to approximate
equation~(\ref{eq:nofr}) by
\begin{equation}
N_V(r) \approx P_V(r) \int_0^{T_{df,V}} p(\tau) d\tau +
\delta(r) \int_{T_{df,V}}^\infty p(\tau) d\tau
\label{eq:nofr2}
\end{equation}
where $P_V(r)$ is defined as the distribution of
displacements averaged over the time $0\le \Delta t\le T_{df,V}$.
Thus
\begin{equation}
N_V(r) \approx \left(1-e^{-T_{df,V}/t_{\rm merge}}\right)P_V(r)
+ e^{-T_{df,V}/t_{\rm merge}} \delta(r)
\label{eq:nofr3}
\end{equation}
and the cumulative distribution describing displacements less
than $r$ is
\begin{equation}
N_V(<r) \approx \left(1-e^{-T_{df,V}/t_{\rm merge}}\right)P_V(<r)
+ e^{-T_{df,V}/t_{\rm merge}}.
\label{eq:nofr4}
\end{equation}

\begin{figure*}
\resizebox{\hsize}{!}
{\includegraphics[scale=0.75,angle=-90.]{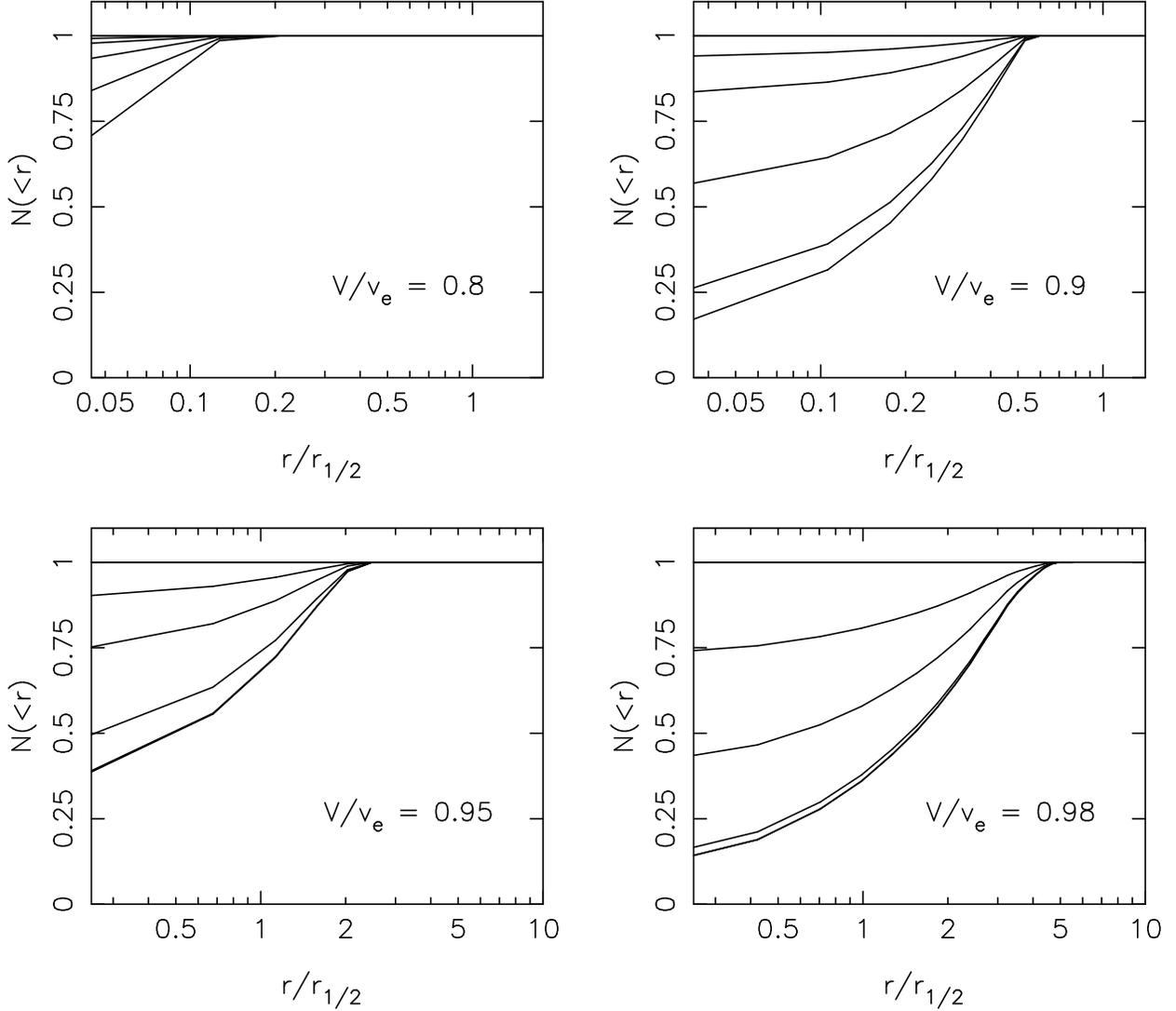}}
\caption{Cumulative distributions $N_V(<r)$ for the
model 1a, for four different values of $V/v_e$.
Each frame plots equation~(\ref{eq:nofr4}) for four
different values of $T_{1/2}/t_{merge}$, where
$T_{1/2}$ is the period of a circular orbit at the
half-mass radius in the equivalent spherical model,
and $t_{merge}$ is the mean time between galaxy mergers,
i.e. between kicks.
The values of $T_{1/2}/t_{merge}$ are
$(0,0.01,0.03,0.1,0.3,1)$; when this ratio is zero,
the cumulative distribution is a step-function, i.e.
the BH would have returned to the origin.
\label{fig:p1}
}
\end{figure*}

\begin{figure*}
\resizebox{\hsize}{!}
{\includegraphics[scale=0.75,angle=-90.]{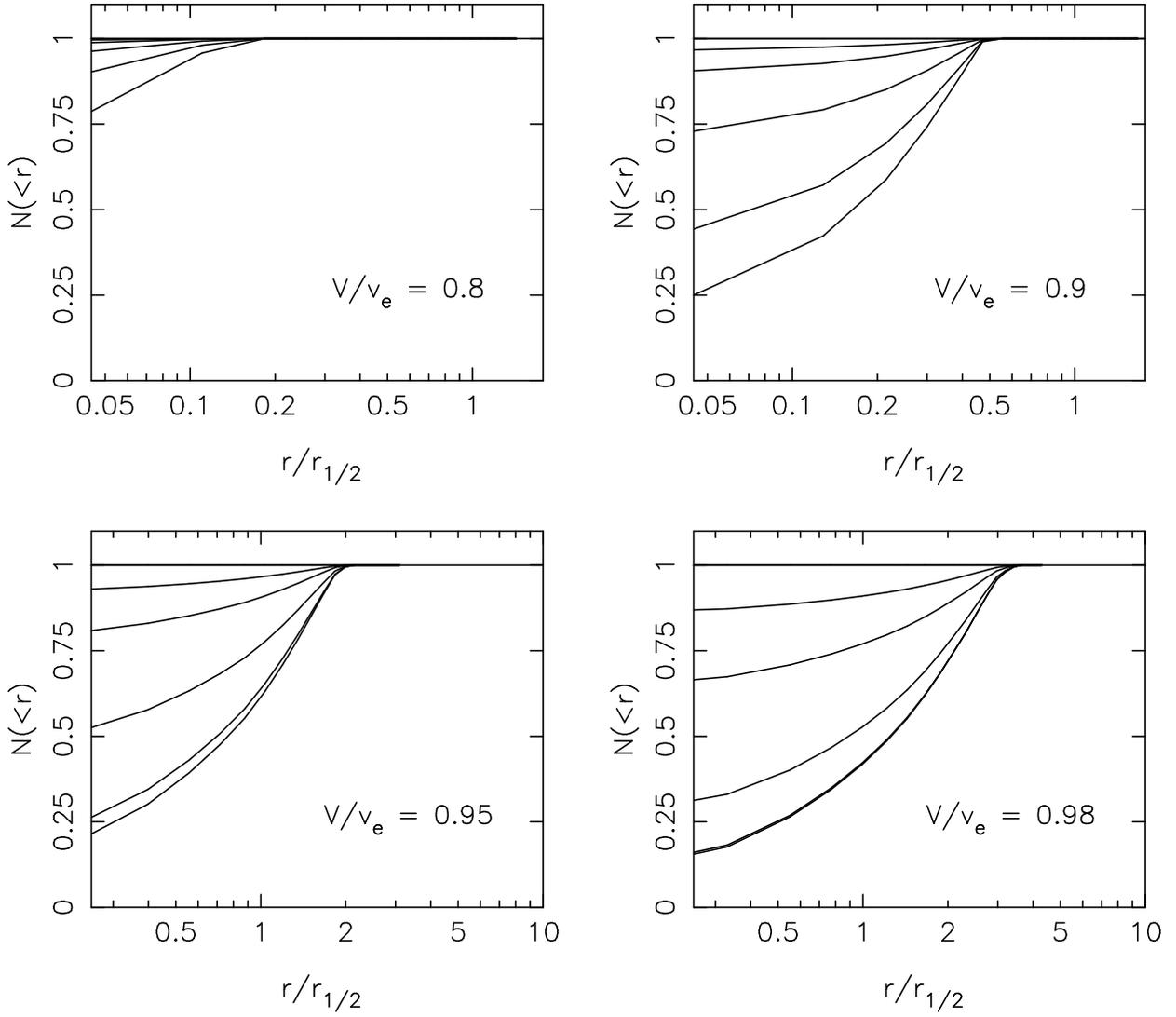}}
\caption{Like Fig.~\ref{fig:p1} but for model 2a.
\label{fig:p2}
}
\end{figure*}

We computed $P_V(<r)$ on a grid of $V/v_e$-values for
several of the models,
then used these numbers to compute the probability
distributions $N_V(<r)$ as a function of the two
parameters
\begin{equation}
\left(T_{1/2}/t_{\rm merge},V/v_e\right)
\end{equation}
where $T_{1/2}$ is a proper reference time defined, as above (Table~\ref{modgal}),
as the period of a circular orbit at the half-mass radius
($\sim r_{1/2}$) in the equivalent spherical model.
Figures~\ref{fig:p1} and~\ref{fig:p2} show the results
for Models 1a and 2a; these models both have $\gamma=1$
and no ``core'' (Table \ref{modincen}), differing only in their
triaxiality (Table \ref{modgal}).
These figures demonstrate that the probability of
finding a kicked BH at a distance $\sim r_{1/2}$
from the center of a galaxy is low unless
the kick velocity is high, $\gap 0.8 v_e$,
and the mean time between mergers is not too long
compared with the crossing time.
Based on Figs. \ref{metvsv1}-\ref{metvsv3c}, this conclusion
would be even stronger
for models like 3b or 3c which have higher central densities.
\clearpage

\clearpage
The largest value expected for $V$ is $\sim 200$ km s$^{-1}$
in the absence of spin (Gonz\'alez et al. 2006;
Sopuerta, Yunes \& Laguna 2006).
By comparison, escape velocities from giant elliptical
galaxies ($M_V\lap -18$) are always greater than
$\sim 400$ km s$^{-1}$ (Merritt et al. 2004).
Hence the probability of finding a significant displaced
supermassive BH in a bright E galaxy is very small, unless a
merger occurred very recently.
Escape velocities from dwarf elliptical (dE) galaxies
($-12\lap M_V\lap -18$) are $\lap 150$ km s$^{-1}$
if only the stellar contribution to the potential
is considered, and $\lap 300$ km s$^{-1}$ if the dark
matter potential is included (Merritt et al. 2004).
(Escape velocities from luminous E galaxies are dominated
by the stars.)
However in these galaxies the mean time between mergers
is believed to be very long, again implying a low
probability of finding a displaced BH.
It is, anyway, worth noting that BH return times would be lengthened
in galaxies containing DM halos. Actually, in luminous galaxies,
central escape velocities are determined essentially by the
stellar distribution; the DM halos have only a small effect
(Merritt et al. 2004).  On the other side, escape velocities in
dE galaxies can be significantly affected by DM, as mentioned in the final
paragraph of previous Section 4. At this regard, useful contribution
will come by the quantitative study of dynamical friction in self-consistent models
of triaxial cuspy galaxies with dark matter haloes, as those obtained by
Capuzzo-Dolcetta et al. (2007) using the Schwarzschild's (1979) orbital superpoasition technique.
Preliminary results are those of a significant
difference in the decay times for energies sufficiently high to
appreciate the role of the (spatially extended) dark matter distribution.
\par As argued above, a BH kicked with velocity $V$
will carry with it material that was orbiting
with velocity $v\gap V$ before the kick.
The size of the region containing this mass is
(Eq. 28)
\begin{mathletters}
\begin{eqnarray}
r_{eff} &\approx& {Gm_\bullet\over V^2}
\approx r_{BH}\left({\sigma\over v_e}\right)^2
\left({v_e\over V}\right)^2 \\
&\approx& 1\ {\rm pc}\ M_8 \sigma^{-2}_{200}
\left({v_e\over V}\right)^{2}
\label{eq:reff}
\end{eqnarray}
\end{mathletters}
with $M_8\equiv M_\bullet/10^8 M_\odot$ and
$\sigma_{200}\equiv \sigma/200$ km s$^{-1}$;
we have set $v_e\approx 4\sigma$, appropriate
for the center of a galaxy.
A substantial displacement requires $V\approx v_e$,
hence the sphere of entrained mass will be of
order a parsec in radius.
This is sufficient to include the
inner accretion disk and much of the
broad-line region gas, implying that a kicked BH
can continue shining for some time as a ``naked''
quasar.
This interpretation has in fact been advanced
for the quasar associated with the HE0450-2958
system, which appears to lack a luminous host
galaxy (Magain et al. 2005; Haehnelt et al. 2005;
Hoffman \& Loeb 2006).
However a number of arguments suggest that the
kick hypothesis is unlikely (Merritt et al. 2005);
for instance, the quasar spectrum exhibits lines
associated with the narrow emission line region
at distances too great to have remain bound to
an ejected BH.
A recent re-analysis of the quasar image
(Kim et al. 2006) also suggests that the presence of a
host galaxy can not be ruled out.
Nevertheless, detection of a broad emission line
spectrum from gas that is displaced spatially or
kinematically from the center
of a galaxy would be strong evidence for a kick,
particularly if the host galaxy exhibited additional
signs of a recent merger.

\section{Conclusions}

We integrated the motion of ``kicked'' BHs in triaxial
models of galaxies, using the expressions derived by
Pesce, Capuzzo-Dolcetta \& Vietri (1992) and
Capuzzo-Dolcetta \& Vicari (2006) for the dynamical friction force 
in an anisotropic stellar distribution.
The velocity dispersion components were computed from
fully self-consistent triaxial models, constructed
via orbital superposition.
We considered different possible forms for the stellar
density at the center of the galaxy, since ejection
of the BH would significantly affect the distribution
of stars there.
Our main results can be summarized as follows:

1. Dynamical friction increases the effective escape velocity
from a galaxy.
This effect is modest, roughly a few percent,
in galaxy models with shallow central density profiles,
but can be very significant in galaxies with $\rho\sim r^{-2}$
central density profiles, since the frictional force acting
on the BH is so strong near the center (Table~\ref{modgal}).

2. Since the dynamical friction force in a triaxial galaxy
depends on angle as well as distance from the center,
escape velocities are a function of ``launching angle'',
being greatest in the direction of the long axis.
Again, this effect is modest in models with low central
concentration but can be appreciable in galaxies with
high central densities (Fig.~\ref{fig:veff}).

3. The time for a kicked BH to return to the center with
zero velocity is longer in a triaxial galaxy than in a
spherical galaxy with the same radial density profile and
length scale $(abc)^{1/3}$, and when kicked with the same
fraction of the escape velocity.
The main reason is that trajectories in the triaxial geometry
are not linear (unless they are exactly along the coordinate axes)
and a kicked BH does not return precisely
through the center, thus reducing the average dynamical
friction force (Figs.~\ref{fig:ex2}, \ref{fig:ex3}).
Infall times are typically several times longer than in the
spherical geometry (Figs.~\ref{metvsv1}-\ref{metvsv3c}).

4. In spite of the delaying effects of the triaxial geometry,
BHs with masses similar to those observed in real galaxies
($M\approx 10^{-3} M_{gal}$)
return to the center in less than $\sim$ a galaxy crossing time,
unless the kick velocity is a large fraction of the escape
velocity.
Since escape velocities in giant elliptical galaxies are large
compared with the maximum kick velocities so far computed
by numerical relativists,
the chance of finding a BH substantially displaced
from the center of such a galaxy is small (Figs. 13, 14).
Escape velocities are smaller in dE galaxies but the mean time
between mergers is probably long, again implying a small probability
of finding a displaced BH.

\acknowledgements
DM acknowledges support from grants AST-0206031, AST-0420920 and AST-0437519 
from the NSF, grant NNG04GJ48G from NASA, and grant HST-AR-09519.01-A from STScI.
This work was supported in part by the Center for Advancing the Study of
Cyberinfrastructure at the Rochester Institute of Technology.

\end{document}